# Linear absorption coefficient of *in-plane* graphene on a silicon microring resonator

Heng Cai,[1] Yahui Cheng,[1] He Zhang,[1] Qingzhong Huang,[1] Jinsong Xia,[1] Regis Barille[2] and Yi Wang[1,*]

[1]*Wuhan National Laboratory for Optoelectronics, Huazhong University of Science and Technology, Wuhan, Hubei430074, China*
[2]*Moltech Anjou, Université d'Angers/ CNRS UMR 6200, PRES UNAM, 2 bd Lavoisiers, 49045 ANGERS cedex , France*
*\*ywangwnlo@mail.hust.edu.cn*

**Abstract:** We demonstrate that linear absorption coefficient (LAC)of a graphene-silicon hybrid waveguide (GSHW) is determined by the optical transmission spectra of a graphene coated symmetrically coupled add-drop silicon microring resonator (SC-ADSMR), of which the value is around 0.23 dB/µm. In contrast to the traditional cut-back method, the measured results aren't dependent on the coupling efficiency of the fiber tip and the waveguide. Moreover, precision evaluation of graphene coated silicon microring resonator (SMR) is crucial for the optoelectronic devices targeting for compact footprint and low power consumption.

Graphene is a two-dimensional material [1-3], which exhibits remarkable optoelectronic characteristics, such as ultrahigh carrier mobility at room temperature [4, 5], ultra-broadband absorption [6, 7], controllable band-gap transition [8], and giant Kerr coefficient [9, 10]. Graphene has been integrated on photonic integrated circuits (PICs) in which the hexagonal carbon sheet is evanescently coupled to the waveguide, leading to unprecedented optical performances[11-16].

Recently, grapheme resting on silicon-on-insulator (SOI) platform offers great potential for optoelectronic devices. Broadband optical modulator [17, 18], mode locked laser[19], broadband photodetectors [20, 21], and enhanced parametric frequency conversion [22] have all been demonstrated utilizing graphene-silicon hybrid waveguides (GSHWs). Owing to its intrinsic physical properties, the linear absorption coefficients (LACs) of GSHWs range from 0.04dB/µm to 0.33 dB/µm [23, 24],which are much larger than those of silicon waveguides [25]. The LACs of GSHWs are dependent on the quality of the transferred graphene and the waveguide configurations [26], thus are required to be optimized for practical applications. For instance, Strong LAC is preferred in electro-absorption modulator (EAM) [27] to obtain a large modulation depth, while in the nonlinear optical application [28], the LAC should be as small as possible to avoid linear absorption. Therefore, it is of great importance to precisely evaluate the optical absorption loss induced by grapheme [29].

Previously, the LACs of graphene-comprising waveguides were obtained by a cut-back method [14-16] in which the light-graphene interaction lengths were varied. And this method is subject to the following limitations:

1. In order to measure the LACs induced by graphene, multiple stripe waveguides should be patterned with different lengths of graphene[14-16], of which the processes are burden some, time consuming and even expensive.

2. The coupling efficiencies between the fiber tips and the waveguides as well as waveguide end facets maintain non-uniformity. Moreover, such fiber-to-waveguide coupling losses are often comparable to or even much higher than the propagation losses in the waveguides [27], leading to high uncertainties in loss measurements, as stated in [30].

3. Because of the imperfect coverage of grapheme with cracks, the measured losses in these stripe waveguides deviate from the linear fit when the lengths of the transferred graphene are beyond 500 µm [14].

In addition to these disadvantages, the silicon microring resonator (SMR), which is a fundamental building block in silicon photonics, has been widely used in integrated optoelectronic devices owing to its compactness of footprint and low requirement of power[31]. Integrating graphene on SMR offers great opportunities in high-performance optoelectronic devices.

In this paper, we present a comprehensive method to determine the LAC of GSHW with a configuration of microring resonator using a small fraction of graphene, which suffers less cracks resulting from grapheme transfer process. A symmetrical add-drop SMR is introduced as a host for graphene. The monolayer grapheme grown by chemical vapor deposition (CVD) is transferred onto the fabricated SMR. The patterning of graphene is accomplished by an electron beam lithography (EBL) process following by a"lift-off" process. The experimental results are compared with a simulation model utilizing Finite Difference Method (FDM).

## 2. Experimental details

The symmetrically coupled add-drop silicon microring resonator (SC-ADSMR) consists of a SMR and dual stripe waveguides working as through and drop ports. The SMR, with a width of 500 nm and a height of 220nm,has a radius of R=15 µm, which is fabricated on a SOI wafer with a buried oxide (BOX) layer of 3 µm using EBL and inductively coupled plasmon (ICP) etching. The stripe waveguides, which are laterally coupled with the SMR, has the same dimensions with that of the SMR and the corresponding gaps are 125 nm and 117 nm respectively. The nearly identical gaps show that the fabricated add-drop silicon microring resonator is symmetrically coupled.

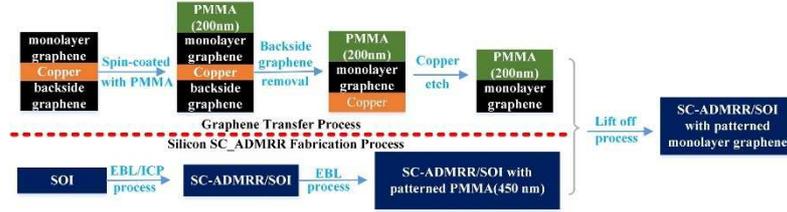

Fig. 1. The processes of the monolayer graphenepreparation and the SC-ADSMR device fabrication.

In Fig. 1, the monolayer graphene transfer and the SC-ADSMR device fabrication processesare illustrated. CVDgrapheneis grown on both sides of the copper foil (Cu). Next, a 200-nmthick poly (methyl methacrylate) (PMMA)(Allresist, AR-P 672.045) film is spin-coated onto the upper surface of the layered backside graphene/Cu/monolayer graphene to protect the monolayer graphene from damage. The sample consisting of layered backside graphene/Cu/monolayer graphene/PMMA is floated on marble's etchant with 15 grams of copper(II) sulfate pentahydrate ($CuSO_4·5H_2O$), 50 ml of deionized water (DI), and 50 ml of concentrated hydrochloric acid (HCl). The backside graphene is totally removed after 2-min chemical etching. Then the Cu/monolayer graphene/PMMA sample is floated on marble's etchant with 15 grams of copper(II) sulfate pentahydrate ($CuSO_4·5H_2O$), 50 ml of deionized water (DI), and 50 ml of concentrated hydrochloric acid (HCl). The backside graphene is totally removed after 2-min chemical etching. Then the Cu/monolayer graphene/PMMA sample is floated on DI for about 3 minutes to remove the chemical impurities. Another marble's etchant with the same chemical recipe is applied for removal of the Cu beneath the monolayer graphene, of which the etching time is about 1.5 hours. The remained sample consisting of layered monolayer graphene/PMMA is last rinsed in DI, leaving it ready for transferring. We employ the following processes to transfer the monolayer graphene to specific region of the SC-ADSMR by ripping out the abundant monolayer graphene: a 450-nm thick PMMA film is spin-coated on the SOI wafer containing the fabricated SC-ADSMR device and the specific region with a 45 degree sector area is patterned using EBL, followed by developing and fixing of the PMMA photo resist. At the time, the prepared monolayer graphene/PMMA sample is transferred onto the chip followed by drying with a mild nitrogen blow. Subsequently, a "lift-off" process is introduced to form the graphene coated SC-ADSMR device. The whole chip is baked at 180°C for 15minutes, after which the PMMA is dissolved in hot acetone for about 1hour. The removal of the PMMA also results in cutting-away of the monolayer graphene resting on the450-nm thick PMMA, which is similar to the lift-off [32] process. Finally, the chip is rinsed in isopropyl alcohol (IPA) for further cleaning and dried again by a mild nitrogen blow. A scanning electron micrograph (SEM) picture of the graphene coated SC-ADSMR device with partial part of the stripe waveguides is shown in Fig. 2(a), in which secondary-electron contrast is obtained. The possible reason for imaging contrast lies in different secondary-electron-generation efficiencies of silicon, silicon oxide and monolayer graphene. At the boundary of the patterned graphene, the random roughness of the monolayer graphene is large due to the coarse-controlled "lift-off" process.

The detailed SEM picture of our fabricated GSHW is shown in Fig. 2(b). At the top of the waveguide, the monolayer graphene is covered. In addition to that, the monolayer grapheme naturally extends to the surface of the BOX layer, forming a trapezoid configuration. Figure 2(c) shows the photonic crystal grating coupler, which is optimized for TE mode coupling exhibiting an insertion loss of 7.5dB per facet.

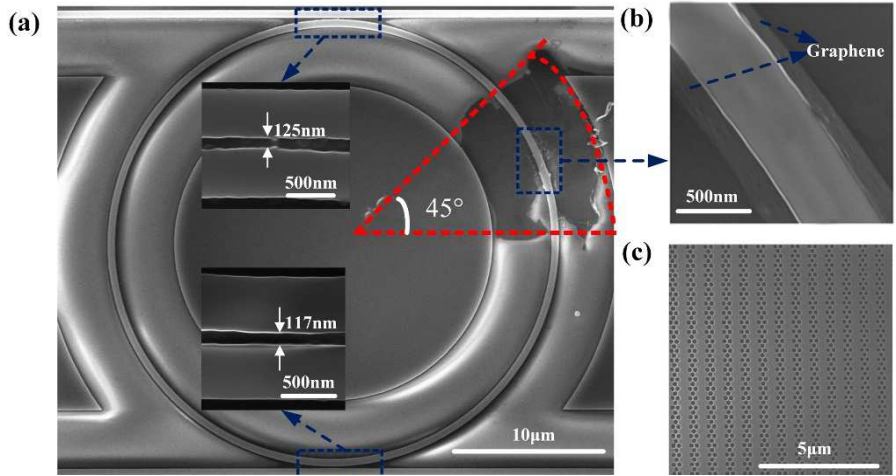

Fig. 2. (a) SEMimage of the SC-ADSMR covered with patterned monolayer graphene (top view). Close-up view of the two coupling regions are also shown. (b) Zoom-in SEM images of the local monolayer graphene covered area in (a), showing that the monolayer graphene uniformly clings to the silicon waveguide. (c) Photonic crystal grating coupler used for coupling light in and out of the SC-ADSMR device.

Adopting the experimental setup in Fig. 3(a), the SC-ADSMR is characterized by an amplified spontaneous emission (ASE) light source (Amonics, AEDFA-300-B-FA) ranging from 1535nm to 1565nm, and the transmission spectra of the through port before and after graphene transfer (noted as woGr and wGr respectively) are recorded by an optical spectrum analyzer (OSA, YOKOGAWA AQ6370),which is shown inFig. 3(b). It shows that the resonator before and after graphene transfer have an identical free spectral range (FSR) of 5.8 nm. The significant reduction of extinction ratio and broadening of the resonances in Fig. 3(b) are the result of increased propagation loss (absorption) induced by the monolayer graphene. The red shift of the resonances are probably caused by the residual PMMA and the chemical impurities introduced by the graphene transfer and "lift off" procedure (the residual PMMA does not attribute to additional loss [24]).

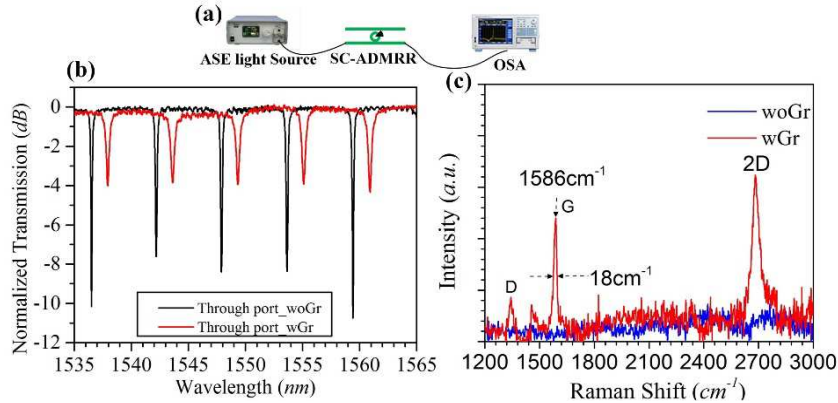

Fig. 3. (a) Experimental setup for characterizing the LAC of GSHW. (b) Opticaltransmission spectra of the SC-ADSMRdevice before and after graphene transfer. (c) Raman spectra measured in the graphene-covered region (red) and in the area where grapheneis lifted off (blue).

The explicit characterization of monolayer graphene relies on the Raman spectrum. In Fig.3(c), we measure the Raman spectra of the graphene coated SC-ADSMR device using LabRAMHR800 (France, Jobin-Yvon). The blue curve represents the Raman spectrum of the area where graphene is lifted off, of which the typical Raman peaks are missing. And the red curve shows a G peak (~ 1586cm$^{-1}$) with a full width at half maximum (FWHM)of~ 18cm$^{-1}$and a 2D peak(~2700 cm$^{-1}$), of which the 2D-to-G peak intensity ratio is about 1.2, implying that the transferred graphene is a monolayer and the corresponding chemical potential is around 0.2 eV [33]. Moreover, a weak D peak is also found at ~1350 cm$^{-1}$, indicating the transferred monolayer graphene is of high quality. We also measure the Raman spectra in some other regions. The results turn out to be the same with the Raman spectra of Fig. 3(c).

## 3. Experimental Results

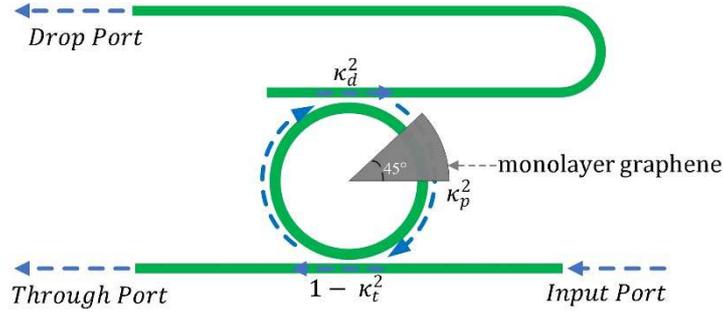

Fig. 4. Scheme of theSC-ADSMR with patternedmonolayer graphene

We adopt an analytical model proposed by Shijun Xiao et al.[30] to characterize the propagation loss in our graphene covered SC-ADMMR device. In Fig 4, the schematic diagram of the model is presented. $\kappa_t^2$ and $\kappa_d^2$ are coefficients representing the fraction of the optical power coupled into and out of the SMR through the input and drop port respectively. $\kappa_p^2$ is the fraction of power losses per round-trip in the SMR due to intrinsic losses mainly resulting from grapheme absorption, roughness induced scattering. In our case, $\kappa_t = \kappa_d$ is satisfied and optical response of the through port is given as [30]:

$$T_{through}(\lambda) = \frac{(\lambda - \lambda_0)^2 + \left(\frac{FSR}{4\pi}\right)^2 (\kappa_p^2)^2}{(\lambda - \lambda_0)^2 + \left(\frac{FWHM_t}{2}\right)^2}, \qquad (1)$$

Where $T_{through}(\lambda)$ is the power transmission of the through port, $FWHM_t$ is the FWHM of the optical transmission spectrum at the through port and $\lambda_0$ is the resonant wavelength. When the input wavelength is at resonance ($\lambda = \lambda_0$), $T_{through}(\lambda)$ has a minimal value of $\gamma_t$ which satisfies:

$$ER_t(dB) = -10\log_{10}(\gamma_t), \qquad (2)$$

With $ER_t$ and $\gamma_t$ the extinction ratio and the minimal power transmission at the through port respectively. The coefficients $\kappa_p$, $\kappa_t$ and $\kappa_d$ are described as:

$$\kappa_p^2 = \frac{2\pi FWHM_t \sqrt{\gamma_t}}{FSR}, \qquad (3)$$

$$\kappa_t^2 = \kappa_e^2 = \frac{\pi FWHM_t \left(1 - \sqrt{\gamma_t}\right)}{FSR}. \qquad (4)$$

The loss per round-trip in the resonatoris determined by:

$$\alpha(dB/round) = -10\log_{10}(1 - \kappa_p^2). \qquad (5)$$

In the proposed sample, an eighth part of the resonator is covered with monolayer graphene. The total losses of the graphene coated SMR includes the linear absorption (LA) of the GSHW and the LA of the silicon waveguide. Assuming that the graphene "lift off" processdoes not induce extra LA in the silicon waveguide, the LAC of the GSHW can be written as:

$$LAC\_GSHW = \frac{(\alpha_{wGr} - \alpha_{woGr}(1-n))}{2\pi nR}. \qquad (6)$$

where $\alpha_{wGr}$ and $\alpha_{woGr}$ correspond to the loss per round-trip before and after graphene transfer respectively, and n=1/8 corresponds to the fractional coverage length ofthe monolayer graphene.

Table 1 shows the detailed experimental results before and after graphene transfer of the fabricated SC-ADMMR device deduced by the transmission spectra in Fig. 3(b), including the resonant wavelengths of the SC-ADSMR before and after graphene transfer, the corresponding extinction ratio and the FWHM of the transmission spectra. Utilizing the analytical method, the LAC of the GSHW in our case has a mean value of 0.23 dB/µm, which is comparable to the results in literature[23, 24].

**Table 1. Parameters for LAC deduction before and after graphene transfer**

| Parameters before and after graphene transfer | | | Results |
|---|---|---|---|
| res_woGr/res_wGr (nm) | ER$_t$_woGr/ER$_t$_wGr (dB) | FWHM$_t$_woGr/FWHM$_t$_wGr (nm) | LAC_GSHW (dB/µm) |
| 1536.52/1537.94 | 9.9/3.8 | 0.170/0.375 | 0.21 |
| 1542.19/1543.62 | 8.2/3.6 | 0.225/0.433 | 0.24 |
| 1547.90/1549.34 | 8.8/3.7 | 0.221/0.405 | 0.22 |
| 1553.65/1555.10 | 9.0/4.0 | 0.231/0.450 | 0.24 |
| 1559.42/1560.90 | 10.8/4.2 | 0.207/0.424 | 0.23 |

## 4. Simulation Model

Graphene can be treated electromagnetically through its surface dynamic conductivity in a complex form consisting of interband and intraband contributions [34]:

$$\sigma(\omega,\mu_c,\tau,T) = \sigma_{intra}(\omega,\mu_c,\tau,T) + \sigma_{inter}(\omega,\mu_c,\tau,T). \quad (7)$$

The intraband contribution can be evaluated as:

$$\sigma_{intra}(\omega,\mu_c,\tau,T) = i\frac{e^2 k_B T}{\pi\hbar^2(\omega+i\tau^{-1})}\left[\frac{|\mu_c|}{k_B T} + 2\ln\left(\exp\left(-\frac{|\mu_c|}{k_B T}\right)+1\right)\right], \quad (8)$$

for $\hbar\omega$ and $|\mu_c| \gg k_B T$, the interband contribution can be expressed as:

$$\sigma_{inter}(\omega,\mu_c,\tau,T) = i\frac{e^2}{4\pi\hbar}\ln\left[\frac{2|\mu_c|-\hbar(\omega+i\tau^{-1})}{2|\mu_c|+\hbar(\omega+i\tau^{-1})}\right], \quad (9)$$

where e is the electron charge, $k_B$ is the Boltzmann constant, $\hbar$ is the reduced Planck's constant, ω is the angular frequency of incident light, T is the temperature, $\mu_c$ is the chemical potential, and τ is the momentum relaxation, respectively. In the simulation, we use the values for the incident wavelength of $\lambda_s$=1550 nm, the temperature of T=300K, and the momentum relaxation of τ=12 fs[35]. The surface dynamic conductivity of graphene σ (ω, $\mu_c$, τ, T) is plotted as a function of the chemical potential $\mu_c$ in Fig. 5.

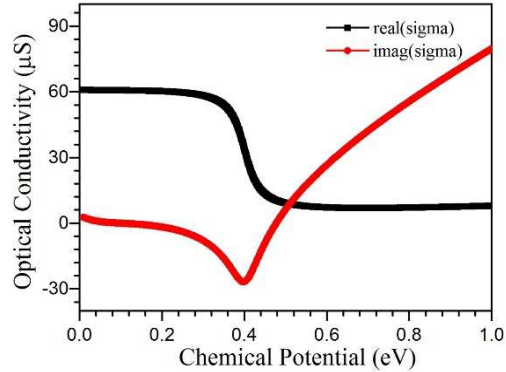

Fig. 5. Surface dynamic conductivity of monolayer graphene as a function of its chemical potential

Taking the monolayer graphene as a boundary surface can be a good approximation in the simulation. We model graphene as a 2D anisotropic boundary and employ the 2D FDM to simulate the propagation loss in a stripe waveguide in Fig. 6(a). Since the graphene is not closely attached to the sidewall of the stripe waveguide, the interaction between the bilateral graphene and light is ignored. The physical boundary condition at the graphene/waveguide interface is expressed as[36]:

$$\mathbf{n}_{12} \times (\mathbf{H}_{2//} - \mathbf{H}_{1//}) = \mathbf{J}_s = \sigma(\omega,\mu_c,\tau,T)\mathbf{E}_{//}, \quad (10)$$

where $\mathbf{n}_{12}$ is the vector normal to the interface, with direction from medium 1 to medium 2, **H** and **E** is the normalized magnetic and electrical field at the interface respectively, $\mathbf{J}_s$ is the surface current density of graphene, and here subscript // denotes *in-plane* field component. Adopting the waveguide width of 500 nm and height of 220 nm, the corresponding losses of the GSHW are plotted with different chemical potentials, of which the value is 0.07dB/μm for TE mode when the chemical potential of graphene is around 0.2eV (Fig. 6(b)).

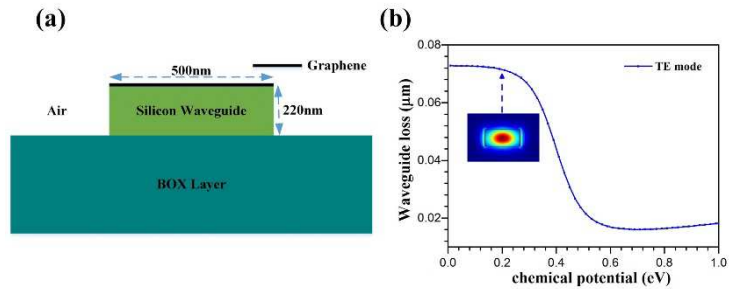

Fig. 6. (a) Cross section of the simulated GSHW (width 500nm, height 220nm). (b) The variation of the LACs of the GSHW as a function of the chemical potential of monolayer graphene. The inset shows the corresponding TE-mode profile of the GSHW with a chemical potential of 0.2 eV.

The simulated LAC of the GSHW is lower than the experimental result. This deviation probably lies in the following reasons:

1. The commercially available chip-sized monolayer graphene grown by CVD has some unpredictablemulti-layer domains [37], which could result in extra loss.
2. Chemical impurities are left after the wet etching transfer of monolayer graphene[38]. These chemical impuritieswill give rise to the surface roughness of the GSHW, leading to extra scattering loss.Moreover, the chemical impurities themselves may contribute to additional absorptionloss.

**5. Conclusion**

We propose a comprehensive way towards direct determination of the LACs of *in-plane*monolayer grapheneintegrated with SC-ADSMR device, whichdoes not depend on the coupling efficienciesbetween the fiber tips and the waveguides as well as the waveguide end facets. We design and fabricate a SC-ADSMR with patterned monolayer graphene.Themeasured result is larger than that of the simulation, which attributes to the imperfection of monolayer graphene and the scattering loss of the abundant chemical impurities introduced by graphene transfer and "lift off" procedures. Our workprovide efficient method to evaluate the linear optical performances of high-performance graphene-comprising waveguides targeting for compact footprint and low power consumption PICs.

**Acknowledgements**

This work is partially supported by National Basic Research Program of China (Grant No. 2012CB922103 and 2013CB933303), and National Scientific Founding of China (Grant No. 60806016 and 61177049). We thank all the engineers in the Center of Micro-Fabrication and Characterization (CMCF) of Wuhan NationalLaboratory for Optoelectronics (WNLO) for the support in device fabrication.